\documentstyle[psfig]{mn}

\newif\ifAMStwofonts

\ifoldfss
  \ifCUPmtlplainloaded \else
    \NewTextAlphabet{textbfit} {cmbxti10} {}
    \NewTextAlphabet{textbfss} {cmssbx10} {}
    \NewMathAlphabet{mathbfit} {cmbxti10} {} 
    \NewMathAlphabet{mathbfss} {cmssbx10} {} 
  \fi
  \ifAMStwofonts
    \ifCUPmtlplainloaded \else
      \NewSymbolFont{upmath} {eurm10}
      \NewSymbolFont{AMSa} {msam10}
      \NewMathSymbol{\upi}     {0}{upmath}{19}
      \NewMathSymbol{\umu}     {0}{upmath}{16}
      \NewMathSymbol{\upartial}{0}{upmath}{40}
      \NewMathSymbol{\leqslant}{3}{AMSa}{36}
      \NewMathSymbol{\geqslant}{3}{AMSa}{3E}

      \let\leq=\leqslant \let\le=\leqslant
       \let\ge=\geqslant
    \fi
  \fi
\fi 

\ifnfssone
  \newmathalphabet{\mathit}
  \addtoversion{normal}{\mathit}{cmr}{m}{it}
  \addtoversion{bold}{\mathit}{cmr}{bx}{it}
  \newmathalphabet{\mathbfit} 
  \addtoversion{normal}{\mathbfit}{cmr}{bx}{it}
  \addtoversion{bold}{\mathbfit}{cmr}{bx}{it}
  \newmathalphabet{\mathbfss} 
  \addtoversion{normal}{\mathbfss}{cmss}{bx}{n}
  \addtoversion{bold}{\mathbfss}{cmss}{bx}{n}
  \ifAMStwofonts
    \ifCUPmtlplainloaded \else
      \UseAMStwoboldmath
      \makeatletter
      \new@mathgroup\upmath@group
      \define@mathgroup\mv@normal\upmath@group{eur}{m}{n}
      \define@mathgroup\mv@bold\upmath@group{eur}{b}{n}
      \edef\UPM{\hexnumber\upmath@group}
      \new@mathgroup\amsa@group
      \define@mathgroup\mv@normal\amsa@group{msa}{m}{n}
      \define@mathgroup\mv@bold\amsa@group{msa}{m}{n}
      \edef\AMSa{\hexnumber\amsa@group}
      \makeatother
      \mathchardef\upi="0\UPM19
      \mathchardef\umu="0\UPM16
      \mathchardef\upartial="0\UPM40
      \mathchardef\leqslant="3\AMSa36
      \mathchardef\geqslant="3\AMSa3E

      \let\leq=\leqslant \let\le=\leqslant
       \let\ge=\geqslant
    \fi
  \fi
\fi 

\ifnfsstwo
  \DeclareMathAlphabet{\mathbfit}{OT1}{cmr}{bx}{it}
  \SetMathAlphabet\mathbfit{bold}{OT1}{cmr}{bx}{it}
  \DeclareMathAlphabet{\mathbfss}{OT1}{cmss}{bx}{n}
  \SetMathAlphabet\mathbfss{bold}{OT1}{cmss}{bx}{n}
  \ifAMStwofonts
    \ifCUPmtlplainloaded \else
      \DeclareSymbolFont{UPM}{U}{eur}{m}{n}
      \SetSymbolFont{UPM}{bold}{U}{eur}{b}{n}
      \DeclareSymbolFont{AMSa}{U}{msa}{m}{n}
      \DeclareMathSymbol{\upi}{0}{UPM}{"19}
      \DeclareMathSymbol{\umu}{0}{UPM}{"16}
      \DeclareMathSymbol{\upartial}{0}{UPM}{"40}
      \DeclareMathSymbol{\leqslant}{3}{AMSa}{"36}
      \DeclareMathSymbol{\geqslant}{3}{AMSa}{"3E}

      \let\leq=\leqslant \let\le=\leqslant
       \let\ge=\geqslant
    \fi
  \fi
\fi 

\ifCUPmtlplainloaded \else
  \ifAMStwofonts \else 
    \def\upi{\pi}
    \def\umu{\mu}
    \def\upartial{\partial}
  \fi
\fi

\title{A FIRST-APM-SDSS survey for high-redshift radio QSOs}
\author[R. Carballo et al]
{R. Carballo,$^1$
\thanks{E-mail:carballor@unican.es},
  J.I. Gonz\'alez-Serrano,$^2$
  F.M. Montenegro-Montes,$^3$
 C.R. Benn,$^4$
\newauthor
K.-H. Mack,$^{3}$ M. Pedani,$^5$ M. Vigotti$^3$
\\
 $^1$ Dpto. de Matem\'atica Aplicada y Ciencias de la Computaci\'on, Univ. de  Cantabria\\
       ETS Ingenieros de Caminos, Canales y Puertos, Avda de los Castros s/n,
 E-39005 Santander, Spain\\
        $^2$ Instituto de F\'\i sica de Cantabria (CSIC-Universidad de
Cantabria),  Avda de los Castros s/n,
 E-39005 Santander, Spain\\
     $^3$ Istituto di Radioastronomia, INAF, via Gobetti 101, I-40129 Bologna,
     Italy\\
$^4$Isaac Newton Group, Apartado 321, E-38700 Santa Cruz de La
Palma, Spain\\
$^5$Centro Galileo Galilei, E-38700 Santa Cruz de La Palma, Spain \\
}

\pagerange{\pageref{firstpage}--\pageref{lastpage}} \pubyear{2003}

\begin{document}

\maketitle

\label{firstpage}

\begin{abstract}
We selected from the VLA FIRST survey a sample of 94 objects with
starlike counterparts in the Sloan Digital Sky Survey, and with APM
POSS-I colour $O-E \ge$ 2, i.e. consistent with their being
high-redshift quasars.  78 of the 94 candidates can be classified
spectroscopically on the basis of either published data (mainly SDSS)
or the observations presented here.  The fractions of QSOs (51 out of
78) and redshift $z >$ 3 QSOs (23 out of 78, 29 per cent) are comparable to
those found in other photometric searches for high-redshift QSOs.  We
confirm that selecting colour $O-E \ge 2$ ensures inclusion of all
QSOs with $3.7 \le z \le 4.4$. The fraction of $2
\le z \le 4.4$ QSOs with broad absorption lines (BALs)
is $27 \pm 10$ per cent (7/26) and the estimated BAL fraction for radio
loud QSOs is at least as high as for optically selected QSOs ($\sim$
13 per cent).  Both the high BAL fraction and the high fraction of
LoBALs among BALs (four to five out of 7) in our sample, compared to previous
work, are likely due to the red colour selection $O-E \ge 2$. The
space density of radio loud QSOs in the range $3.7 \le z \le 4.4$
($\bar{z} = 4.0$) with $M_{\rm AB} (1450) \le -26.6$ and $P \ge
10^{25.7}$ W Hz$^{-1}$ is $ 1.7 \pm 0.6$ Gpc$^{-3}$.  Adopting a
radio-loud fraction of $13.4 \pm 3$ per cent, this corresponds
to $\rho = 12.5 \pm 5.6$ Gpc$^{-3}$, in substantial
agreement with the cumulative luminosity function of SDSS QSOs in
Fan et al. (2001a). We note the unusual flat-spectrum radio-luminous
QSO FIRST 1413+4505 ($z=3.11$), which shows strong associated
Ly$\alpha$ absorption (rest-frame equivalent width $\sim 40$ \AA ) and
an extreme observed luminosity, $L \sim 2 \times 10^{15} L_\odot$.

\end{abstract}


\section{Introduction}

This is the third of a series of papers presenting new samples of
high-redshift radio QSOs selected by matching the FIRST catalogue of
radio sources (Faint Images of the Radio Sky at Twenty-cm; Becker,
White \& Helfand 1995, White et al. 1997) with red starlike objects
from the APM (Automated Plate Measuring Facility) catalogue of the
POSS-I survey (McMahon \& Irwin 1992).

Papers I and II (Benn et al. 2002, Holt et al. 2004) reported a
sample of 18 $z>3.6$ QSOs including the largest sample of $z>4$
radio-selected QSOs then known.  The search was carried out within
a $\sim$ 7030 deg$^2$ region using the constraints: i) $E \leq
18.8$ and starlike, ii) $S_{\rm 1.4 \ GHz} \ge 1$ mJy, and iii)
colour selection $O - E \ge 3$. This colour range includes an
estimated $95 \pm 1.5$ per cent of $E <$ 18.8 QSOs with redshift
3.8 $< z <$ 4.5 (Vigotti et al. 2003).
On the basis of the 13 QSOs with $z \simeq 3.8 - 4.5$ and
comparing to an equivalent sample at $z \simeq$ 2 drawn from the
FIRST Bright Quasar Survey of the north Galactic cap (FBQS-2,
White et al. 2000), we showed (Vigotti et al. 2003) that the
decline in space density of $M_{\rm AB}$ (1450 \AA) $\le $ $-$26.9
QSOs ($H_\circ = 50$ km s$^{-1}$ Mpc$^{-1}$, $\Omega_M=1$ and
$\Omega_\Lambda=0$ here and throughout the paper) was
approximately a factor 2 between $z \sim 2$ and $z \sim 4$,
significantly smaller than the value $\sim$ 10 found for samples
including lower luminosity objects (Fan et al. 2001a).

The Sloan Digital Sky Survey Data Release 3 (SDSS, Stoughton et
al. 2002; DR3 Abazajian et al. 2005) provides moderately deep CCD
imaging in five bands $ugriz$ covering $\sim$ 5282 deg$^2$.  We
present and discuss a new sample of high-redshift radio-selected
QSO candidates in a 1378.5 deg$^2$ area of overlap between FIRST
and SDSS DR3 in the north Galactic cap.  The selection criteria
are: i) $E \le 19.1$ and starlike in APM, ii) $S_{\rm 1.4 \ GHz}
\ge 1$ mJy, iii) radio-optical separation less than 1.5 arcsec,
iv) colour $O - E \ge 2$ (including $O$ non-detections) and v)
starlike in SDSS.  This new sample, with a wider colour range, has
several advantages. Firstly, the SDSS photometric catalogue
provides reliable morphological classification of the sources,
allowing us to readily eliminate the galaxies classed as starlike
in APM POSS-I.
Secondly, SDSS
provides $\sim$ 3 \AA-resolution spectra and spectroscopic
classifications of many objects, particularly those selected as
QSO candidates on the basis of their $ugriz$ colours, or as
counterparts of FIRST sources. We continue to use the APM
catalogue for colour selection since our previous work showed a
high efficiency and completeness in the selection of $z > 3.85$
QSOs using $O-E \ge 3$ and with the new limit $O-E \ge 2$ we can
check that no $z >$ 3.7 QSOs have $O-E <$ 3.

The paper is structured as follows. In Section 2 we present the
sample and the status of the spectroscopic classification. Section
3 reports optical spectroscopy of part of the sample.  The
spectroscopic classification of the sample as QSOs, narrow
emission line galaxies or stars is presented in Section 4.1. The
distribution of optical magnitudes and $O-E$ colours is discussed
in Section 4.2. In Section 4.3 the sample is compared with
previous radio-selected QSO samples from the literature, in terms
of the selection criteria and the resulting QSO redshift
distribution. In Section 4.4 we comment briefly on the spectra of
seven QSOs exhibiting strong blueshifted broad absorption lines
(BALs) and we analyse the fraction of BAL QSOs in the sample.
Section 4.5 is devoted to the peculiar QSO FIRST 1413+4505. In
Section 4.6 we compute the absolute magnitudes, $k$-corrections
and radio luminosities of the 10 QSOs with $z \ge 3.7$, and we
discuss the completeness of a sub-sample of seven of them. In
Section 5 we use this sample to calculate the space density of
QSOs. Section 6 summarizes our conclusions.

\begin{table}
\caption{Right ascension and declination ranges of the four areas of sky included in the survey. }
\begin{tabular}{|c|c|}
\hline
RA (h)& DEC (deg)\\
\hline
  $~7.933 ~\rightarrow ~~8.667 $ & $+30 ~\rightarrow ~+46$ \\
  $~8.733 ~\rightarrow ~~9.600 $ & $+38 ~\rightarrow ~+55$ \\
  $~9.667 ~\rightarrow ~15.667$ & $+42 ~\rightarrow ~+58$ \\
  $15.733 ~\rightarrow ~16.667$ & $+30 ~\rightarrow ~+45$ \\
\hline
\end{tabular}
\end{table}

\begin{figure*} \psfig{figure=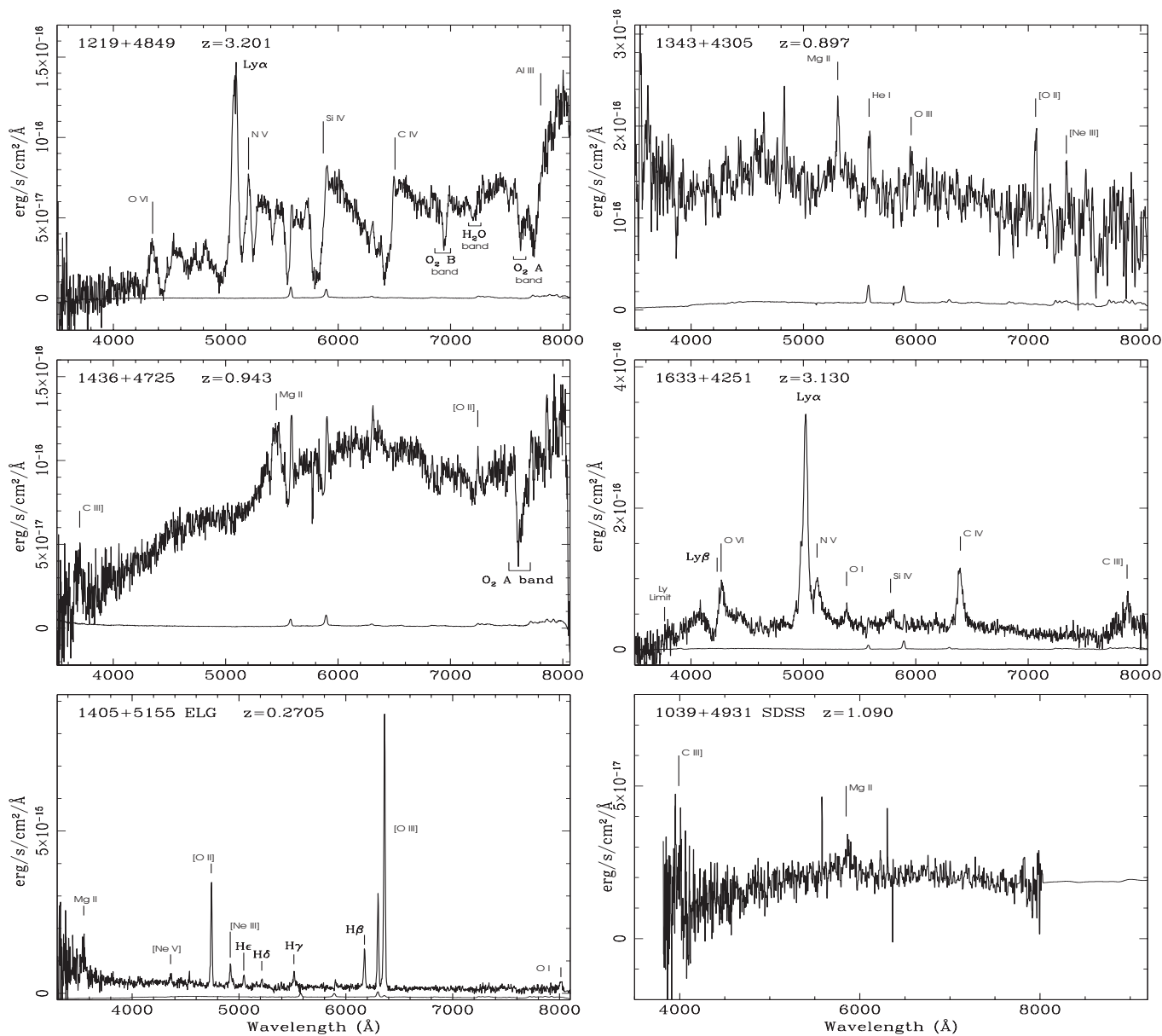,height=16.0cm,width=18cm}
 \caption{TNG spectra of the four QSOs and the emission-line galaxy
identified for the first time here, and the SDSS spectrum of
source FIRST 1039+4931, which was classified `unknown' in SDSS.
The night sky spectrum (scaled) is shown near the zero level in
the TNG spectra. The emission features are labelled by ion
(wavelengths from Vanden Berk et al. 2001). }
\end{figure*}

\begin{figure*}
\psfig{figure=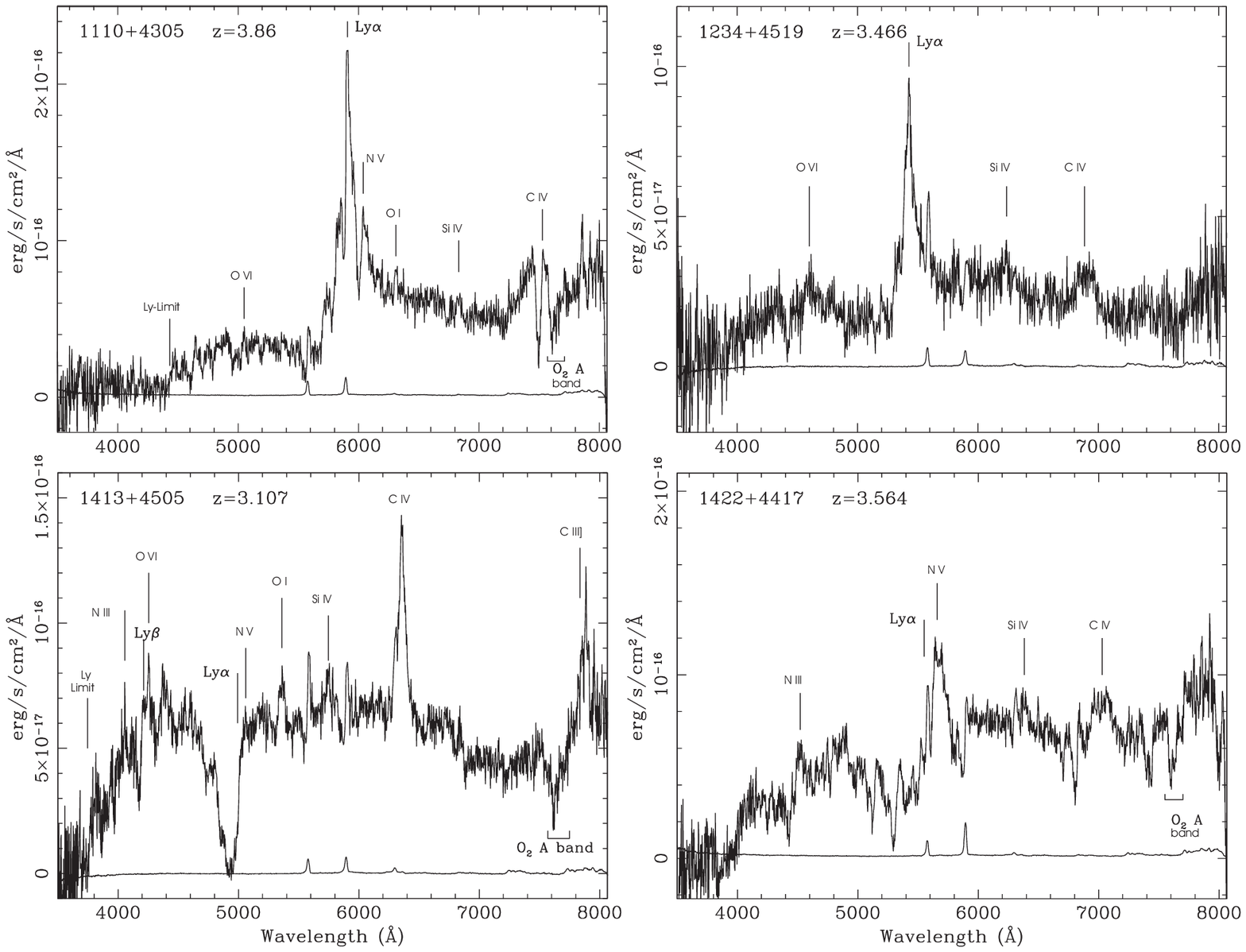,height=11cm,width=18cm} \caption{Spectra
of four QSOs identified at the TNG and also observed by SDSS DR4.}
\end{figure*}

\section{Selection of the sample}

The sample was selected from the 1378.5-deg$^2$ area defined in
Table 1. This area includes most of the region covered by SDSS DR3
in the north Galactic cap, which is also covered by the FIRST
survey and by the APM catalogue of POSS-I.  The FIRST survey
includes 122463 sources in this area with $S_{\rm 1.4 \ GHz}$
(peak) $\ge$ 1 mJy, of which 113 have APM $E <$ 19.1, $O-E >$ 2.0
and lie within 1.5 arcsec of starlike objects in SDSS DR3.
Eighteen of these were undetected in APM POSS-I $O$ but were
detected in APS POSS-I $O$ and had APS($O-E$) $ < 2$, and were
therefore removed from the sample. For the source FIRST 1340+5619
we found a large difference between the SDSS and APM magnitudes
($r=23.51$ versus $E=18.99$) and the source was eliminated from
the sample after confirming with SDSS that the APM counterpart is
a blend.  The final sample thus includes 94 candidate
high-redshift radio QSOs.

Of these 94 candidates, spectra were first obtained for seven in
papers I and II, for six in the literature (found using the NASA
Extragalactic Database - NED) and for 41 by SDSS DR3 (which also
reobserved nine of the thirteen previously discovered).  In
Section 3 we present TNG optical spectroscopy of 13 (randomly
selected) of the 40 remaining candidates, and we classify an SDSS
DR3 source given spectral class `unknown' in the Sloan survey.
Subsequent to these observations, SDSS DR4 (2005 June 30) reported
spectroscopy of 10 of the remaining 27 candidates (and also of
four of those observed here). In total, 78 of the 94 candidates
(83 per cent) are now spectroscopically classified (see Section
4.1).

\section{TNG optical spectroscopy}

Spectra of 11 candidates (indicated in column 9 of Table 2) were
obtained with the Telescopio Nazionale Galileo (TNG) on 2005 March 10
and 11 using the DOLORES (Device Optimized for LOw RESolution)
spectrograph in long-slit mode. The LR-B grism was used, yielding a
wavelength range 3000 -- 8800 \AA\ and dispersion 2.9 \AA
~pixel$^{-1}$. The detector was a thinned, back-illuminated Loral CCD
with 15$\mu$m pixels. Exposure times were typically 900 s. Three
spectrophotometric standard stars were observed in order to calibrate
the instrumental spectral response.  The seeing was $\approx$ 1.8
arcsec, and the width of the slit was set to 2 arcsec, yielding a
spectral resolution of 23 \AA, as measured from sky lines.

Standard data reduction was carried out using the IRAF\footnote{IRAF
is distributed by the National Optical Astronomy Observatories, which
is operated by Association of Universities for Research in Astronomy,
Inc., under cooperative agreement with the National Science
Foundation.} package.  Arc-lamp exposures were used for the wavelength
calibration, and yielded solutions with rms residuals $<$ 1.5 \AA\
. Small spectrum-to-spectrum wavelength shifts were corrected using
sky lines in the science spectra. In most of the spectra, the two or
three brightest sky emission lines were not completely removed in the
sky subtraction.  This did not affect spectral classification or
redshift measurement, but for presentation in Figs. 1, 2 we manually
cleaned some of the strongest residuals.

Spectra of two more candidates (FIRST 1343+4305 and FIRST 1405+5155)
were obtained with similar instrumental set-up on the nights of 2005
July 12 and 13. Two spectra per object were taken, shifting the object
along the slit to minimize the effect of detector artefacts and remove
the OH sky lines. For each object, one 2D spectrum was subtracted from
the other, to remove the background, then wavelength calibrated,
aligned in the spatial direction and coadded. Exposure times were $2
\times 1800$ s for 1343+4305 and $2 \times 1200$ s for 1405+5155.

The 13 objects include eight QSOs, one $z=0.2705$ emission line galaxy
(ELG), and four late-type stars.  The redshift of each QSO was
estimated as the average of the values measured from individual
emission-line centroids (excluding Ly$\alpha$, which is often affected
by Ly$\alpha$ forest absorption).  Six of the QSOs have $3.1 < z <
3.9$, the other two have $z=0.897$ and $z=0.943$.

The TNG spectra of the eight QSOs and the ELG are shown in Figs. 1
and 2.  Fig. 1 also shows the SDSS DR3 spectrum of FIRST
1039+4931, classified `unknown' in SDSS, but which we identify as
a QSO at $z=1.09$.  Four of the eight QSOs were also observed in
SDSS DR4, published after our TNG observations took place.  The
TNG spectra of these four objects are shown in Fig. 2.

\section{Properties of the sample}

\subsection{Spectroscopic classification}

\begin{figure}
\psfig{figure=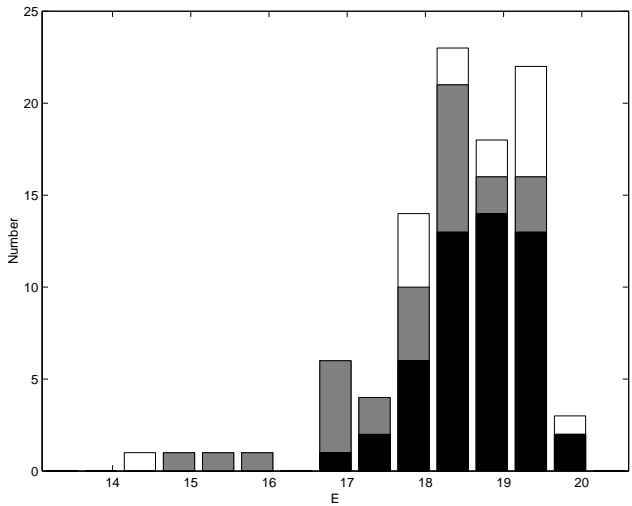,width=8.3cm} \caption{Distribution of
corrected $E$ magnitude. QSO identifications are shown in black,
stars and ELGs in grey and unclassified sources in white.}
\end{figure}

Table 2 lists the basic optical and radio properties of the 94
objects in the sample, together with the spectroscopic
classification and redshift, mostly from SDSS DR3 and DR4. The
spectra were classified as one of: QSO (broad emission lines),
emission-line galaxy (narrow emission lines), late-type star or
early-type star.  The sample of 94 objects includes 51 QSOs (0.27
$< z <$ 4.31), 25 stars, two ELGs ($z=0.27, 0.31$) and 16 objects
with no optical spectrum, i.e. 78 out of 94 (83 per cent) are
classified.  Of the 78 objects with spectra, 38 per cent (30/78)
are QSOs with $z\ge 2$, 29 per cent (23/78) are QSOs with $z\ge 3$
and 13 per cent (10/78) are QSOs with $z\ge 3.7$, confirming the
efficiency of the adopted selection criteria for identifying
moderate- to high-$z$ QSOs.  The efficiencies of previous radio
QSO surveys are compared in Section 4.3.

\begin{figure*}
\psfig{figure=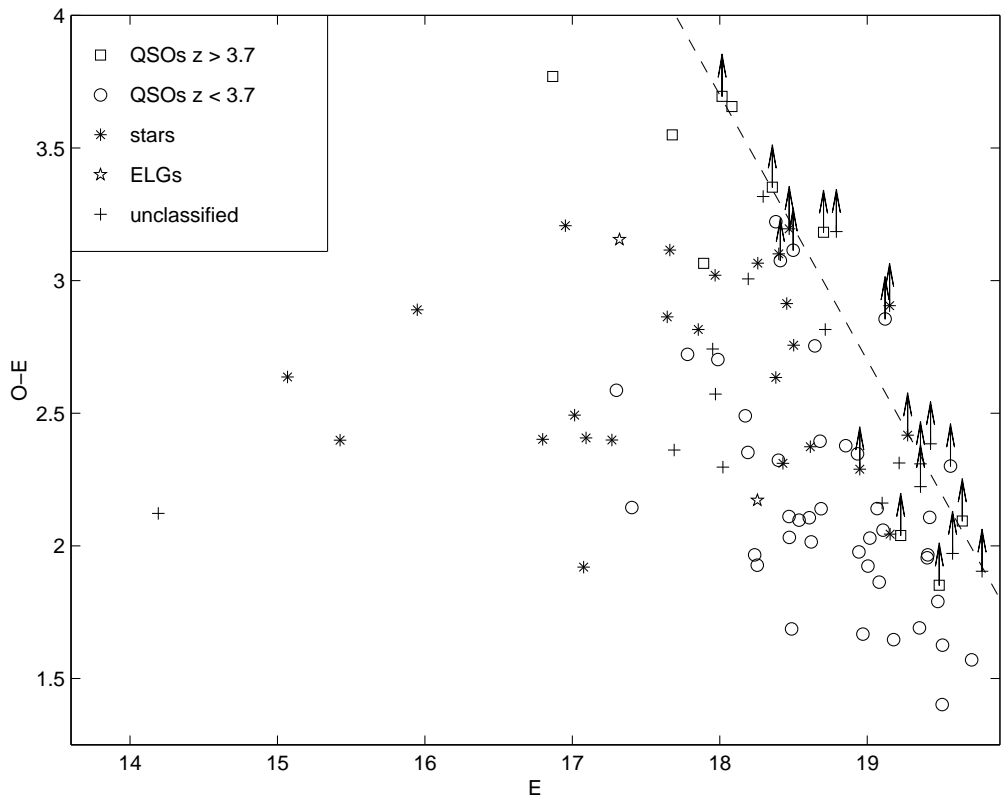,width=13.2cm} \caption{Colour-magnitude
diagram for all 94 candidates in the sample. The dashed line
corresponds to an assumed POSS-I blue plate limit $O = 21.7$
(referred to the APS calibration). Arrows show objects not
detected in $O$ band.}
\end{figure*}

\subsection{Distribution of optical magnitudes and $O-E$ colours}

In Table 2 and hereafter, unless otherwise indicated, the quoted
POSS-I APM magnitudes are recalibrated with respect to APS
(McMahon et al. 2002), and are corrected for Galactic reddening.
This results in some of the sample having final $O-E < 2$ and $E >
19.1$ (the recalibrated $E$ magnitudes are typically $0.3 \pm 0.3$
mags fainter than the original APM values). The histogram of the
$E$ magnitudes is shown in Fig. 3.

Fig. 4 shows the distribution $O-E$ versus $E$. QSOs have a
broader $O-E$ colour range than stars. The ratio QSOs/stars
increases from 21/12 in the range $17.1 < E \leq 18.6$ to 29/5 for
$E>18.6$ (see also Fig. 3). Assuming that the unclassified sources
in Fig. 4 are either QSOs or stars, the distribution in $O-E$
versus $E$ suggests that $\sim 60$ per cent are QSOs.

Fig. 5 (see also Fig. 6b) shows that the QSO colours redden
significantly with redshift: the four measured values of $O-E$ and
three out of six lower limits imply $O-E >$ 3 for $z >$ 3.7. This
is due to the drop in intensity in the $O$ band as it becomes more
dominated by the region blueward of the Lyman limit, and the
increase in intensity in the $E$ band as Ly$\alpha$ enters the
band.

\setcounter{table}{2}
\begin{table}
 \centering
 \begin{minipage}{80mm}
 \caption{Selection criteria and QSO fractions for
FBQS-2 and for the current sample}
\begin{tabular}{|l|r|r}
\hline
                              &FBQS-2         & This work            \\
\hline
Area (deg$^2$)                & 2682~~~      & 1378.5~~~             \\
$E$                           &$\le 17.8$~~~      & $\le 19.1^{(1)}$ \\
$O-E$                         & $\le 2$~~~   & $\ge 2^{(1)}$         \\
Radio-opt separation (arcsec) & $\le 1.2$~~~ & $\le 1.5$~~~          \\
Starlike morphology           & $E$ or $O$~~ & $E$ and SDSS          \\
                              &              &                       \\
Number of candidates          &1238~~~       & 94~~~                \\
Number of spec. class.        &1130~~~       &  78~~~                \\
Number of QSOs                 &636~~~        &  51~~~                \\
Number of QSOs $2 \le z < 3$   &53~~~         &   7~~~                \\
Number of QSOs $3 \le z < 3.7$ &12~~~         &  13~~~                \\
Number of QSOs $z \ge 3.7$     &0~~~          &  10~~~                \\
\hline
\end{tabular}

1: $O$ and $E$ magnitudes as taken directly from APM

\end{minipage}
\end{table}

\begin{figure}
\psfig{figure=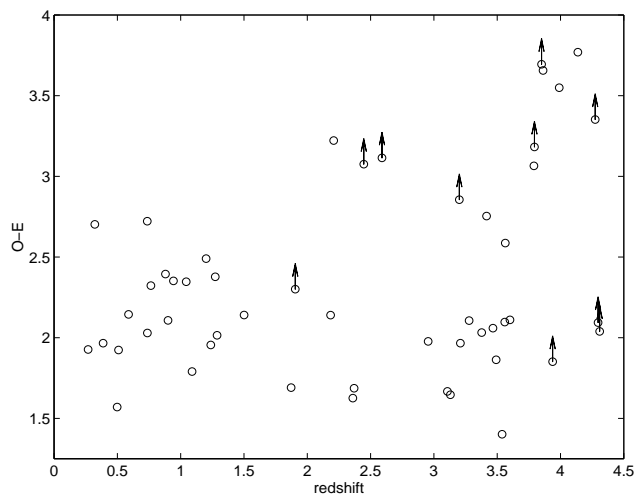,width=8.3cm} \caption{  $O-E$ versus
redshift for the QSOs in the sample.}
\end{figure}

An important conclusion from Fig. 5 is that the colour cut $O-E
\ge 2$ includes most of the QSOs with $3.7 \le z < 4.4$, and this
result is in agreement with previous studies. In Vigotti et al.
(2003) we estimated a fraction of $z>3.8$ QSOs with APM$(O-E)>3$
of $95 \pm 1.5$ per cent, using a sample with $E$(APM) $\le 18.8$.
In FBQS-2 (White et al. 2000, $O-E \le 2$ and $E \le 17.8$) 90 per
cent of the QSO candidates were spectroscopically classified, and
these included 636 QSOs, none of which had $z>3.5$, although 12
had $3 < z <3.5$. We are therefore confident that the number of
QSOs we find with $3.7 \le z \le 4.4$ is limited only by the radio
and $E$ band flux density limits, with no QSOs missed due to the
colour selection.

\begin{table*}
 \begin{minipage}{160mm}
  \caption{
Well-defined samples of $z \ge 3$ radio-selected QSOs}
\begin{tabular}{|l|c|c|c|c|c|}
\hline
Reference, radio/opt limits& Area  (deg$^2$)      &Cands.&Spec. Class.&$z>3.0$ &  $z>3.7$\\
(1)&(2)&(3)&(4)&(5)&(6)\\
\hline
Hook et al. 1998                                      &1600        & 73& 50~~&~~6/50 ~~12\% &~~0/50 ~~~0\% \\
Green Bank $S_{\rm 5 ~GHz} \ge 25$ mJy                &            &   &   &              &              \\
FIRST $S_{\rm1. 4 ~GHz} ~~~\alpha_{1.4}^{5} \ge -0.5$ &            &   &   &              &              \\
APM POSS-I $E\le 19.5$,  ~$O-E\ge 1.2$                &            &   &   &              &              \\
                                                      &            &   &   &              &              \\
Snellen et al. 2001                                   &6400        & 27& 27~~&~~9/27 ~~33\% &~~5/27 ~~19\% \\
Green Bank $S_{\rm 5 ~GHz} \ge 30$ mJy                &            &   &   &              &              \\
NVSS $S_{\rm 1.4 ~GHz} ~~~\alpha_{1.4}^{5} \ge -0.35$ &            &   &   &              &              \\
APM $E<19$, $O-E>2$                                   &            &   &   &              &              \\
                                                      &            &   &   &              &              \\
Hook et al. 2002                                      &7265 $BR$   &228&202~~&~18/202 ~~9\% &~12/202 ~~6\% \\
Green Bank $S_{\rm 5 ~GHz} \ge 50$ mJy                &(3637 $BRI$)&   &   &              &              \\
NVSS $S_{\rm 1.4 ~ GHz} ~~~ \alpha_{1.4}^{5} \ge -0.5$&            &   &   &              &              \\
APM UKST $R<21$; $B-R\ge 1.5$                         &            &   &   &              &              \\
or $R-I\ge 1$ or $B-I\ge 2$                           &            &   &   &              &              \\
                                                      &            &   &   &              &              \\
Holt et al. 2004                                      &7030        &194&121~~&~15/121 ~12\% &~15/121 ~12\% \\
FIRST $S_{\rm 1.4 ~ GHz} \ge 1$ mJy                   &            &   &   &              &              \\
APM  $E<18.8 $, ~$O-E>3$                              &            &   &   &              &              \\
                                                      &            &   &   &              &              \\
This work                                             &1380        & 94& 78~~&~23/78 ~~29\% &~10/78 ~~13\% \\
FIRST $S_{\rm 1.4 ~ GHz} \ge 1$ mJy                   &            &   &   &              &              \\
APM $E \le 19.1$, ~$O-E \ge 2$                        &            &   &   &              &              \\
\hline
\end{tabular}

Col. (1):  reference, optical and radio catalogues, and optical
and radio flux-density and spectral-index ($S_\nu \propto
\nu^\alpha$) limits. (2): survey area. (3,4,5,6): number of
candidates,  number of candidates with spectroscopic
classification and the fraction of the latter at redshifts $z > 3$
and $z>3.7$.

\end{minipage}
\end{table*}

\subsection{Comparison with other radio samples}

\begin{figure}
\psfig{figure=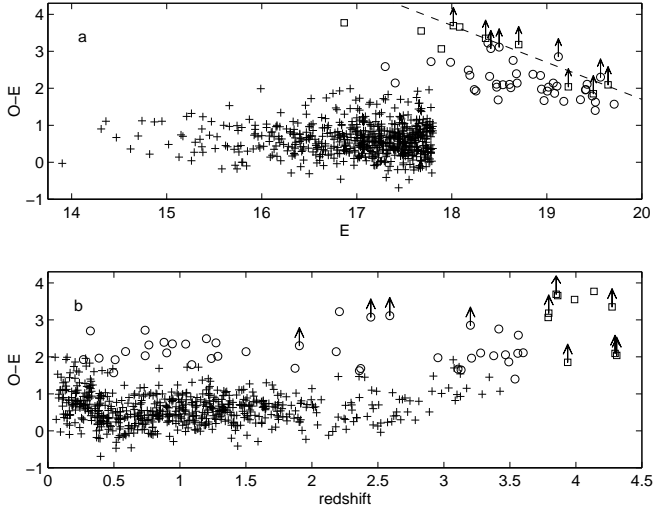,width=8.6cm} \caption{(a) $O-E$ versus
$E$ for the QSOs in FBQS-2 (crosses) and in this work (circles for
$z < 3.7$, squares for $z \ge 3.7$).  The dashed line corresponds
to the assumed POSS-I blue plate limit $O$ = 21.7.(b) $O-E$ versus
$z$ for the same QSOs.
}
\end{figure}

The survey reported here can be considered complementary to
FBQS-2.  Both surveys select FIRST radio sources with starlike
optical counterparts, but the colour criteria are $O-E \le 2$ for
FBQS-2 (blue-excess objects) and $O-E \ge 2$ for our work.  In
FBQS-2 objects were selected with $O-E \le 2$ to eliminate
galaxies, since catalogues based on photographic plates provide
poor discrimination between stellar and non-stellar objects.  The
survey selection criteria, and the QSO fractions as a function of
redshift, are compared in Table 3. Figs. 6a and 6b show $O-E$
versus $E$ and $O-E$ versus $z$ for the QSOs in the two surveys.
Whereas the redshifts covered by FBQS-2 range up to 3.4, our
sample includes redshifts up to 4.3 (Fig. 6b, Table 3), and this
is a consequence of both the colour cut and the fainter
magnitudes.

The efficiency of QSO selection in this work (51/78 = 65 per cent)
is slightly larger than in FBQS-2 (636/1130 = 56 per cent).
Although our selection criteria filter out most of the QSOs below
$z=3.5$, and these are covered by FBQS-2, our survey is better at
rejecting galaxies, since we take the morphological information
from SDSS whereas FBQS-2 used the less reliable APM
discrimination. Although our sample has fewer sources, the
efficiency for moderate to high redshift QSOs is high: 20/78 = 26
per cent for $2 \le z \le 3.7$ compared to 65/1130 = 6 per cent
for FBQS-2. At $z \ge 3.7$ the efficiency of our selection is
10/78 = 13 per cent.  The maximum redshift in FBQS-2 is $z=3.4$.

Table 4 compares the sizes and search efficiencies of various
radio-selected high-redshift QSO samples. The success-rate for
high-$z$ QSOs and the number of QSOs in this work are comparable to
or larger than those of previous searches for radio-selected QSOs at
much {\it higher} radio flux densities.

 \subsection{The fraction of BAL QSOs}

\begin{figure*} \psfig{figure=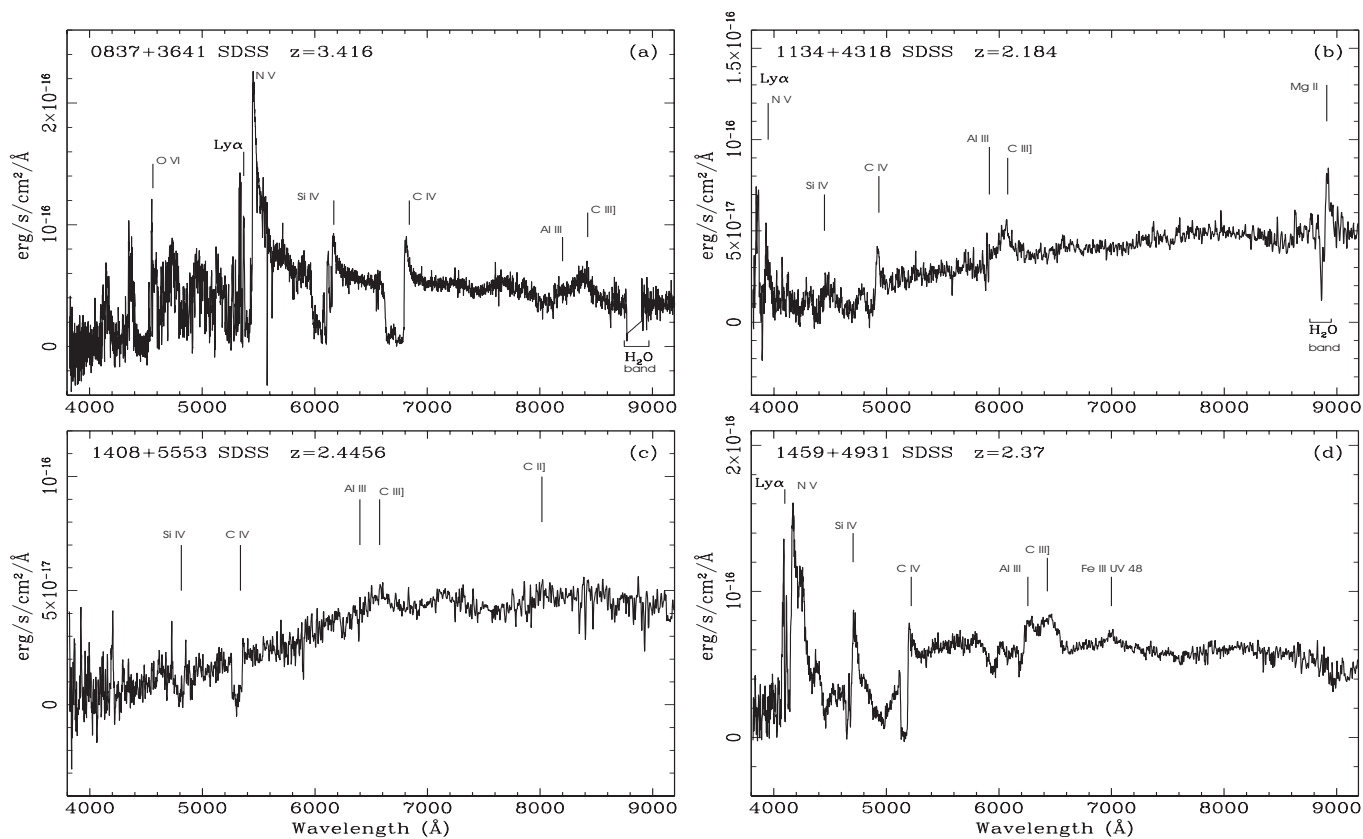,height=11.0cm,width=18cm}
 \caption{SDSS spectra of four of the seven $2 \le z \le 4.4$ BAL QSOs
 in the current sample. The three remaining spectra are presented in
 Fig. 1 of this paper, in Benn et al. (2002) and in Benn et
 al. (2005).}
\end{figure*}

Twenty-six of the QSOs with spectra have redshifts $ 2.0 \le z \le
4.4$, so that Si {\sc iv} 1397 \AA~ and C {\sc iv} 1549 \AA~ are
included in the observed wavelength range.  Seven of these QSOs show
strong broad absorption lines (BALs) with velocity widths above 2000
km s$^{-1}$ and reaching velocities above 3000 km s$^{-1}$,
and were classified as BALs. The balnicity index (BI) of each BAL was
computed following the prescription of Weymann et al. (1991) and is
presented in Table 5.

The fraction of BALs in our sample is therefore 7/26 = $27 \pm 10$
per cent. For comparison, from the study of the BAL QSOs in FBQS-2
(Becker et al. 2000) we derive an observed fraction of BALs for
$1.5 \le z \le 3$ and $15 \le E \le 17.8$ of 14/134 = $10 \pm 3$
per cent (BALs with BI=0 excluded). In view of the redder colours
found for BALs, compared to non-BALs, and the adopted colour
selection, $O-E \le 2$, Becker et al. conclude that the BAL
fractions derived from their work should be considered as lower
limits. Our result, $27 \pm 10$ per cent, confirms a larger
fraction of BALs in a red-selected sample ($O-E \ge 2$), and
following their argument this value should be considered as an
upper limit to the actual fraction. We note however, that the
distribution of balnicity indices for the FBQS-2 BALs reaches
lower values than the one in Table 5, which is more similar to the
BI distribution of the BAL sample in Weyman et al. (1991),
nowadays regarded as highly conservative. In particular we have
6/7 cases with BI $\ge$ 4500 km s$^{-1}$ (86 per cent), Weyman et
al.  24/40 (60 per cent) and Becker et al. 6/14 (43 per cent).
Because of our selection of the stronger BALs, in the comparison
with FBQS-2 the fraction $27 \pm 10$ per cent should be regarded
as a lower limit. Differences in the optical magnitudes and
redshift ranges, along with small number statistics could explain
the differences in the BAL fractions between FBQS-2, $>10 \pm 3$
per cent, and our work, $ 27 \pm 10$ per cent.

Reichard et al. (2003a) obtained for a QSO sample selected from
SDSS Early Data Release  at $1.7 \le z \le 4.2$ (regardless of
radio emission) an observed BAL fraction $14 \pm 1$ per cent. The
BI distribution of these BALs reaches low values, with only 14/185
= 8 per cent of them having BI $\ge$ 5000 km s$^{-1}$ (Reichard et
al. 2003b). Reichard et al. (2003a) confirm that BAL QSOs are
redder than non-BALs, and estimate a true BAL fraction of $13 \pm
1$ per cent, taking into account colour-dependent selection
effects in BALs and non-BALs. Trump et al. (2006) found from a
larger sample of SDSS DR3 QSOs at $1.7 \le z \le 4.38$ a slightly
lower BAL fraction, $10.4 \pm 0.2$ per cent, with BI $\ge$ 5000 km
s$^{-1}$ for 162/1756 = 9 per cent of the BALs. For the comparison
with our work and FBQS-2, the two last values should be regarded
as upper limits, giving the low balnicity indices reached and the
steeply rising distribution of BI towards low values (Reichard et
al. 2003b).  Taken together, the results from this work, Becker et
al., Reichard et al. and Trump et al. suggest that the BAL
fraction in radio-selected QSO samples is at least as high as that
of optically-selected QSOs.

\begin{table}
 \begin{minipage}{160mm}
  \caption{
Balnicity indices of the $z>2$ BAL QSOs}
\begin{tabular}{|l|l|r|l|c}
\hline
~FIRST & ~$z_{\rm QSO}$&BI ~~~~ &BAL& Spectrum\\
~name  &    &km s$^{-1}$  &type& \\
\hline
0837+3641 &3.416& 5600~~ &LoBAL  &SDSS, Fig. 7\\
1134+4318 &2.184& 9800~~ &LoBAL? &SDSS, Fig. 7\\
1219+4849 &3.201& 4800~~ &LoBAL  &TNG,  Fig. 1\\
1408+5553 &2.446& 4500~~ &       &SDSS, Fig. 7\\
1459+4931 &2.370& 12850~~&LoBAL  &SDSS, Fig. 7\\
1516+4309 &2.59 & 6000~~ &LoBAL  & Benn et al. 2002\\
1624+3758 &3.377&2990~~  &       & Benn et al. 2005\\
\hline
\end{tabular}
\end{minipage}
\end{table}

A brief description of the seven BALs is given below:
\begin{itemize}

\item FIRST 0837+3641 shows C
{\sc iv} and Si {\sc iv} troughs starting at the QSO redshift, with a
velocity width $\sim$ 8500 km s$^{-1}$ and sharp onset. The absorption
bluewards of the expected Al {\sc iii} $\lambda 1860$
emission line suggests that
this QSO is a low-ionization BAL (LoBAL).

\item FIRST 1134+4318 shows C
 {\sc iv} and Si {\sc iv} troughs starting at the QSO redshift. The C
 {\sc iv} absorption presents two clearly distinct components. The
 weak absorption bluewards of the expected Al {\sc iii} emission
 line suggests that this QSO could be a LoBAL.

\item FIRST 1219+4849 shows  C {\sc iv} and Si {\sc iv} troughs, the former starting sharply
at the emission redshift and extending almost to the Si {\sc iv}
emission line.  There is absorption bluewards of the Al {\sc iii}
emission line, so this is probably a LoBAL, although some
contamination by sky bands cannot be excluded.

\item FIRST 1408+5553 shows C {\sc iv}
and Si {\sc iv} troughs of width $\sim$ 5500 km s$^{-1}$, starting
at the emission redshift.

\item FIRST 1459+4931 shows  C {\sc iv}
and Si {\sc iv} troughs with a sharp onset at the emission
redshift. The C {\sc iv} absorption extends to the Si {\sc iv}
emission line and clearly shows two components. The strong Al {\sc
iii} absorption confirms this QSO as a  LoBAL. We identify the
emission line at 7000 \AA~ as Fe {\sc iii} UV 48 (rest frame
wavelengths  2062.2, 2068.9 and 2079.65 \AA, see Laor et al.
1997).

\item FIRST 1516+4309, with deep C {\sc iv} and
Si {\sc iv} troughs was firstly reported as a BAL in Benn et al. (2002).
The absorption bluewards of the expected Al {\sc iii} suggests that
this QSO is a LoBAL.


\item FIRST 1624+3758 is an unusual BAL
(Benn et al. 2005), with a clear detached C {\sc iv} trough extending
from $-21000$ to $-29000$ km s$^{-1}$.

\end{itemize}

The fraction of LoBALs in our sample, four to five out of seven or
55-70 per cent, is unusually high. For comparison, the observed
fractions obtained from the FBQS-2 and from the SDSS BALs in Reichard
et al. (2003a) are $28\pm 14$ per cent (four out of fourteen) and $13
\pm 3$ per cent (24/181, $1.7 \le z \le 3.9$) respectively. The higher
fraction in our sample is consistent with Becker et al. result that
BALs in general and the LoBALs in particular are over-abundant among
the reddest QSOs. Moreover, Reichard et al. (2003b) found that LoBALs
have stronger CIV absorption troughs (i.e., stronger BI) than HiBALs,
and the balnicity indices in our sample are in fact higher than the
typical values in FBQS-2 and in Reichard et al.(2003a) SDSS sample.

\subsection{The peculiar QSO FIRST 1413+4505}

FIRST 1413+4505 (Fig. 2, $z=3.107$) has an unusual spectrum, with
a broad strong absorption starting at the expected position of the
Ly$\alpha$ emission line and extending up to $-18500$ km s$^{-1}$
bluewards. We identify the feature as H {\sc i} on the basis of
the location of the line and the detection of Ly$\beta$ absorption
at this redshift. The broad absorption is not detected in metal
lines. The Ly$\alpha$ absorption feature has two components. The
first one, at $z_{abs}=3.06$, has a rest-frame EW $\sim 38$\AA\
and associated Ly$\beta$ absorption with EW $\sim$ 5\AA. The
second component, at $z_{abs} = 2.93$, has EW $ \sim 7$\AA\ and
the corresponding Ly$\beta$ absorption is difficult to identify
due to the high noise in this part of the spectrum. The EW of the
first Ly$\alpha$ absorber is consistent with a single damped
Ly$\alpha$ system with high H {\sc i} column density, i.e. $>
10^{21}$ cm$^{-2}$ (Lanzetta et al. 1991), at the QSO position.
However, the Ly$\alpha$/Ly$\beta$ ratio suggests an optically thin
absorber, therefore the system could be a velocity-resolved
BAL-like wind. A higher resolution spectrum is needed to determine
the nature of this absorption system.

The spectral energy distribution of the source can be obtained
from NED. The FIRST integrated flux density at 1.4 GHz is $140.3
\pm 0.1$ mJy, and the flux density at 4.85 GHz is $125 \pm 15$ mJy
(Becker, White \& Edwards 1991).  The source is unresolved in
FIRST, $<$ 1 arcsec. The SDSS magnitudes, corrected for Galactic
extinction, are $u=22.32 \pm 0.23$, $g= 20.15 \pm 0.04 $, $r=
19.27 \pm 0.02$, $ i= 19.09 \pm 0.02 $ and $z= 19.09 \pm 0.05$.
The source is not detected in 2MASS and the magnitude limits for
99\% completeness are: $J=15.8$, $H=15.1$ and $K=14.3$ (see 2MASS
Web site\footnote{http://www.ipac.caltech.edu/2mass}). FIRST
1413+4505 is highly luminous in the radio, $P_{\rm 1.4 \ GHz} =
2.2 \times 10^{27}$ W Hz$^{-1}$, and has a flat radio spectral
index, $\alpha=-0.09$. From the $E$ magnitude and using the TNG
spectrum for the $k$-correction we obtained $M_{\rm AB}$ (1450
\AA) = $-25.7$ (see equations in Vigotti et al. 2003).

At the position of the QSO, mid-infrared emission is detected via 1D
addition of IRAS Scans (SCANPI) at 12$\mu$m and 25$\mu$m.  The
automatic procedure obtained via NED gives an 11$\sigma$ detection at
12$\mu$m in the addition modes `detector weighted' and `mean',
15$\sigma$ detection for `rms weighted' and non-detection for the
`median' mode.  The corresponding peak flux densities in Jy are $0.22
\pm 0.02$, $0.21 \pm 0.02$ and $0.31 \pm 0.02$, respectively.
At 25$\mu$m coaddition gives  3.6$\sigma$ detection for `detector weighted' and `mean' modes,
5$\sigma$ for `rms weighted' and non-detection with the `median' mode.
The corresponding peak flux densities are $0.08 \pm 0.02$, $0.08 \pm
0.02$ and $0.11 \pm 0.02$.

\begin{table*}
 \centering
 \begin{minipage}{160mm}
 \caption{Absolute magnitudes, radio luminosities
and contributions to space density, of the $z \ge 3.7$ QSOs}
\begin{tabular}{|c|c|c|c|r|c|c|c|c|c|c|c|}
\hline
 RA    & DEC   & $z$ &      $E$ & $S_{1.4}$&$M_{\rm AB}$ &log ~$P_{\rm 1.4}$ &$z_{\rm min,o}$&$z_{\rm max,o}$& $z_{\rm max,r}$&$z_{\rm a}$ & $\rho$   \\
 J2000 & J2000 &     &          &mJy       &1450 \AA   &  W Hz$^{-1}$        &               &              &  &              & Gpc$^{-3}$\\
(1)&(2)&(3)&(4)&(5)&(6)&(7)&(8)&(9)&(10)&(11)&(12)\\
\hline
 09  41  19.44&  51  19  33.0&    3.850 & 18.02 &   2.49& -27.67   & 26.00&3.70&4.40&  4.40 &  3.70-4.40 & 0.145 \\
 10  57  56.28&  45  55  53.1&    4.137 & 16.87 &   1.10& -28.37   & 25.70&3.70&4.40&  4.38 &  3.70-4.38 & 0.149 \\
 11  02  01.91&  53  39  12.7&    4.297 & 19.64 &   4.45& -25.84   & 26.33&    &    &       &       &       \\
 11  10  55.22&  43  05  10.1&    3.862 & 18.08 &   1.21& -27.41   & 25.69&3.70&4.40&  4.32 &  3.70-4.32 & 0.162 \\
 12  40  54.92&  54  36  52.2&    3.938 & 19.49 &  15.09& -25.85   & 26.80&    &    &       &       &       \\
 13  09  40.70&  57  33  09.9&    4.274 & 18.36 &  11.33& -26.59   & 26.73&3.85&4.40&  4.40 &  3.85-4.40 & 0.187 \\
 15  06  43.81&  53  31  34.4&    3.790 & 17.89 &  14.63& -27.57   & 26.76&3.70&4.40&  4.40 &  3.70-4.40 & 0.145 \\
 15  10  02.93&  57  02  43.4&    4.309 & 19.23 & 254.97& -26.53   & 28.09&    &    &       &       &       \\
 16  19  33.65&  30  21  15.1&    3.794 & 18.70 &   3.88& -26.96   & 26.18&3.70&4.40&  4.40 &  3.70-4.40 & 0.145 \\
 16  39  50.52&  43  40  03.7&    3.990 & 17.68 &  25.23& -27.92   & 27.03&3.70&4.40&  4.40 &  3.70-4.40 & 0.145 \\
\hline
\end{tabular}

Cols. (1, 2, 3, 4, 5): similar to cols. (1, 2, 11, 6, 4) in Table
2. (6): absolute magnitude $M_{\rm AB}$ at rest-frame 1450 \AA.
(7): log radio luminosity at rest-frame 1.4 GHz. (8, 9, 10, 11):
redshift limits in the optical ($z_{\rm min,o}$, $z_{\rm max,o}$),
radio ($z_{\rm max,r}$), or both ($z_{\rm a}$), within the range
$3.7 \le z \le 4.4$. (12): contribution to the space density.

\end{minipage}
\end{table*}

Mid-infrared detection at $z \ge 2$ is rare and we checked if the
detection with SCANPI could arise due to confusion with one of the two bright stars 52
arcsec NE and 48 arcsec SW (binary) of the QSO, but neither of them
was detected at 12$\mu$m or 25$\mu$m.

Assuming that the mid-infrared emission is physically associated
with the QSO, and not a chance coincidence, the luminosity of the
QSO in the rest frame wavelength range 2200 \AA \ -- 6.1 $\mu$m
(from $z$-band to 25 $\mu$m in the observer frame) would be $1.8
\times 10^{15} L_\odot$, i.e. FIRST 1413+4505 would be one of the
most luminous objects known. To our knowledge there are five other
systems with such extreme luminosities, and although they were
discovered in different ways, and have different SEDs and
absorption/emission-line properties, the extreme luminosity was in
all cases ascribed to flux magnification due to gravitational
lensing. These objects are the Seyfert 2 like object IRAS FSC
10214+4724 (Rowan-Robinson et al. 1991), the BAL QSO H1413+117
(Magain et al. 1988), the sub-mm galaxy SMM 02399-0136 (Ivison et
al. 1998), the optically bright QSO APM 08279+5255 (Irwin et al.
1998) and the Lyman break galaxy MS 1512-cB58 (Yee et al. 1996).

\subsection{The $z \ge 3.7$ QSO sample}

Fig. 5 suggests that no QSOs with $z \ge 3.7$ will have been
missed by the colour selection $O-E >$ 2. The properties of the 10
$z >$ 3.7 QSOs are given in Table 6. Adopting the magnitude limit
$E \le 19.1$, leaves 7 QSOs in the range $3.7 \le z \le 4.4$.

The absolute magnitudes at rest-frame 1450-\AA, $M_{\rm AB}$(1450
\AA), of the $z \ge 3.7$ QSOs were calculated from the
extinction-corrected $E$ magnitudes using the same procedure as in
Vigotti et al. (2003), except for a better estimation of the
$k$-corrections, obtained here from the individual spectra rather
than using an average. We used the optical spectra from this work
(FIRST 1110+4305), Benn et al. 2002 (FIRST 0941+5119) or from SDSS
(remaining QSOs).

Fig. 8a shows $m_{\rm AB} [1450 (1+z) ]- E$, or $k$-correction,
for the redshift range 3.7 -- 4.4, using the spectra of the 10
QSOs. Fig. 8b shows the mean and standard deviation of $m_{\rm AB}
[1450 (1+z)] - E$ for the seven QSOs with $E \le 19.1$.

The radio luminosities at rest-frame frequency of 1.4 GHz were
calculated assuming a spectral index $\alpha = -0.3$ ($S_\nu
\propto \nu^\alpha$), which is the median spectral index obtained
for 13 $z >$ 3.6 FIRST-APM QSOs  in Holt et al. (2004).  The radio
luminosities range from $P_{\rm 1.4 ~GHz}=10^{25.7}$ to
$10^{28.1}$ W Hz$^{-1}$, so all ten QSOs are `radio loud' QSOs,
adopting the criterion $P_{\rm 1.4 ~GHz} > 10^{25.5}$ W Hz$^{-1}$
(Gregg et al. 1996).

\begin{figure} \psfig{figure=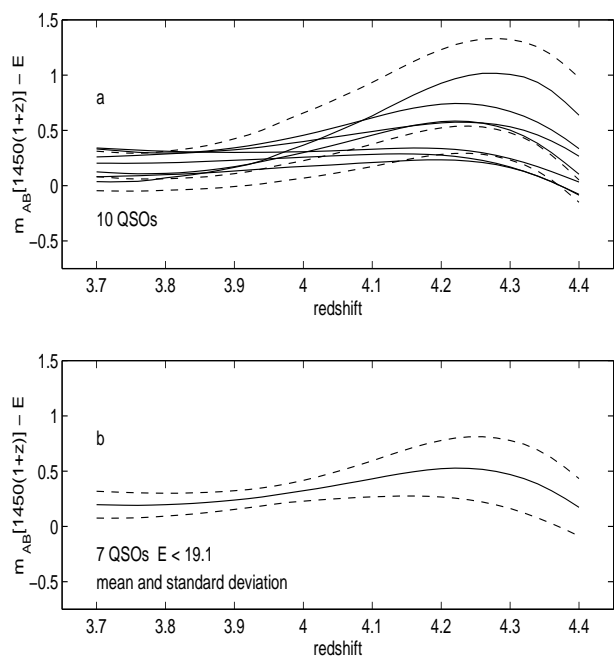,height=8.6cm,width=8cm}
\caption{(a) $k$-correction between the AB magnitude at 1450
$(1+z)$ \AA~ and through the $E$ band, versus redshift,
obtained  from the spectra of the $z \ge 3.7$ QSOs. Solid lines
correspond to the 7 QSOs with $E \le 19.1$. (b) Average
$k$-correction for these 7 QSOs (solid line) and standard
deviation (dashed lines).}
\end{figure}

%

\subsubsection{Completeness of the $z \ge 3.7$, $E \le 19.1$ QSO sub-sample}
There are several sources of incompleteness, summarised below.

(1) Only 78 of the 94 candidates were spectroscopically
classified, giving a fraction of 83 per cent.

(2) The APM completeness for $E \le 19.1$ was estimated as the
fraction of SDSS $r \le 19.8$, $z \ge$ 2 QSOs in the SDSS DR3
Quasar Catalog (Schneider et al. 2005) detected and starlike in
APM $E$. The adopted $r$ band limit was obtained from the average
magnitude difference $r-E=0.69$ (standard deviation 0.34) of the
10 high-$z$ QSOs in the sample. Average $r-E$ and standard
deviation for the 94 candidates are 0.55 and 0.36 respectively.
From the 2060 DR3 QSOs in the surveyed area of this work, 2002
were detected in APM and 1793 of these were starlike, giving a
fraction $87 \pm 2$ per cent.

(3) The completeness of the FIRST catalogue for the range 1 -- 1.5
mJy is $\sim$ 70 per cent (Prandoni et al. 2001). For the sample
of seven QSOs, with radio flux densities: 1.1, 1.2, 2.5, 3.88,
11.33, 14.6 and 25.2 mJy, we estimate a completeness of $92 \pm 5$
per cent, assuming Poisson errors.

(4),(5) In Vigotti et al. (2003) we found for a very similar
sample completeness $98 \pm 1$ per cent due to QSOs probably
missed because of extended radio emission, and 99 per cent
completeness due to QSOs that may exceed the radio-optical
separation limit of 1.5 arcsec.

(6) From our work and from the literature (see Figs. 5, 6 and
Section 4.2) we estimate 100 per cent completeness for the colour
cut $O-E \ge 2$ in the redshift range 3.7 -- 4.4.

(7) The completeness of the SDSS photometric survey quoted in the
project web site, computed by comparing the number of objects
found by SDSS to the number found by the COMBO survey (Classifying
Objects by Medium-Band
Observations\footnote{http://www.mpia-hd.mpg.de/COMBO}), is 100
per cent for point sources with SDSS $r \le 21$, which is above
the maximum $r = 20.6$ in our sample.

The combined
completeness due to the above seven factors is $64 \pm 5$ per cent.

\section{QSO space density at high-redshift}

The observed space density of QSOs in the redshift range $3.7 \le
z \le 4.4$, over the 1378.5 deg$^2$ survey area, was calculated
using the sub-sample of seven QSOs with $E \le 19.1$ and the
$1/V_a$ estimator (see e.g. Avni \& Bahcall 1980). The space
density contributed by each QSO was computed as the inverse of the
available volume, using the constraints $E \le 19.1$, $S_{\rm 1.4
~GHz} \ge 1$ mJy and $3.7 \le z \le 4.4$. The redshift limits and
the space density contribution of each QSO are listed in Table 6.
The sum of these contributions yields a space density (for $M_{\rm
AB}(1450) \le -26.6$) of $1.1 \pm 0.4$ Gpc$^{-3}$, assuming
Poisson errors. The mean redshift of the QSOs is 4.0, the mean
absolute magnitude $M_{\rm AB}(1450) = -27.5$ and the mean log
radio luminosity $P_{\rm 1.4 \ GHz} = 10^{26.3}$ W Hz$^{-1}$.
Correcting for the $64 \pm 5$ per cent completeness, the space
density is $1.7 \pm 0.6$ Gpc$^{-3}$. Assuming the radio-loud
fraction $13.4 \pm 3$ per cent for $P_{\rm 1.4 \ GHz} \ge
10^{25.7}$ W Hz$^{-1}$ from Vigotti et al. (2003), the space
density of all QSOs with $3.7 \le z \le 4.4$ and $M_{\rm AB}
(1450) \le -26.6$ is $12.5 \pm 5.6$ Gpc$^{-3}$.

\subsection{Comparison with other surveys at $z \ge 3.7$}

In Vigotti et al. (2003) we computed $\rho [M_{\rm AB} (1450) \le
-26.9] = 7.4 \pm 2.6$ Gpc$^{-3}$, at  $3.85 \le z \le 4.45$,
assuming the same radio-loud fraction as above. The average
redshift (4.2), optical luminosity ($M_{\rm AB}(1450) = -27.7$)
and radio luminosity ($ P_{\rm 1.4 \ GHz} = 10^{26.6}$ W
Hz$^{-1}$) were all slightly larger than for the current sample.
Although the space density derived from the new sample has a
larger statistical error, this work extends the computation of the
space density to lower optical luminosities and benefits from
better $k$-corrections.

Four of the seven $z \ge 3.7$ QSOs in the current work were also
included in Vigotti et al. The $M_{\rm AB} (1450)$ values obtained
in the two analyses agree within 0.15 magnitudes, except for FIRST
1309+5733, with $M_{\rm AB} (1450)=-26.6$ compared to $-27.1$ in
Vigotti et al. The reason for the discrepancy is the high
$k$-correction for this source (the highest among $E\le 19.1$ QSOs
in Fig. 8).

Fan et al. (2001a,b) obtained the space density of QSOs at high
redshift using an SDSS-selected sample of 39 QSOs  with $3.6 \le z
\le 5.0$ and $-27.75 \le M_{AB}(1450) \le -25.5$. Seven of these
QSOs have redshifts and absolute magnitudes within the range used
here ($z$ = 3.7 -- 4.4, $M_{\rm AB} \le -26.6$) and in this sense
our sample, with seven radio loud QSOs, has comparable quality in terms of the
number of objects.

Fig. 9 shows the $1/V_a$ estimators of the cumulative luminosity
function obtained in this work and in Vigotti et al., along with
the maximum likelihood solution by Fan et al. For $z=4.0$ Fan et
al. give $\rho [M_{\rm AB} (1450) \le -26.6] = 8.0 \pm 4.6$
Gpc$^{-3}$, compared to $12.5 \pm 5.6$ Gpc$^{-3}$ in this work.
These two values are consistent, within the errors.

\begin{figure}
\psfig{figure=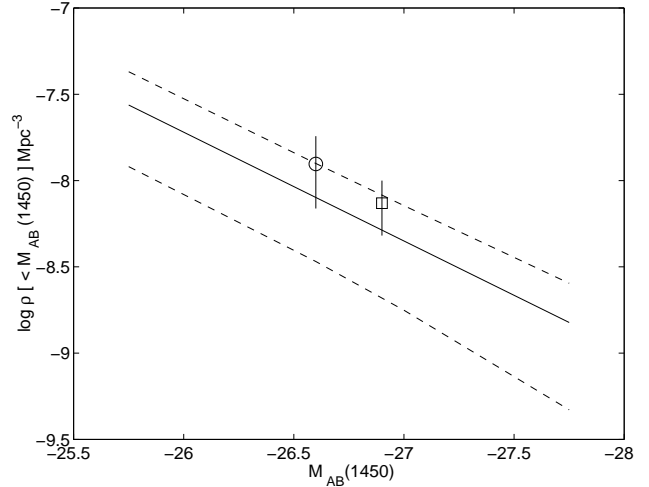,width=8.3cm} \caption{ Cumulative
luminosity function derived from the $1/V_a$ estimator in this
work (circle, $3.7 \le z \le 4.4$) and in Vigotti et al. (square,
$3.8 \le z \le 4.5$). Solid and dashed lines: maximum likelihood
solution and standard deviation from Fan et al. ($3.6 \le z \le
5.0$).}
\end{figure}

\section{Conclusions}

A sample of 94 radio-emitting QSO candidates has been derived by
cross-correlating the FIRST survey, APM POSS-I and the SDSS
photometric survey, selecting objects with $E \le$ 19.1, $O-E \ge
2$, and starlike in SDSS. 78 of the 94 sources (83 per cent) are
spectroscopically classified, mainly from SDSS, but also from the
literature and from TNG spectroscopy presented here (13 sources).
The classified sources include 51 QSOs, with redshifts 0.27 $< z
<$ 4.31. Our main results are as follows:

(i) The efficiency of
selection of high-redshift QSOs is
23/78 = 29 per cent  for $z \ge 3$,
 and 10/78 = 13 per cent for $z \ge 3.7$,
comparable to that for previous searches for high-redshift
QSOs at much {\it higher} radio flux densities.

(ii) We found no 3.7 $< z <$ 4.4 QSOs with 2 $< O-E <$ 3, supporting our
assumption that no $z > 3.7$ QSOs are missed by selecting $O-E \ge$ 2.

(iii) The BAL fraction for the QSOs with $2 \le z \le 4.4$ in our
sample is $\sim 27 \pm 10$ per cent (7/26), larger than found for
FBQS-2 (Becker et al. 2000) and for the optically-selected SDSS
samples in Reichard et al. (2003) and Trump et al. (2006), despite
these three works being more complete for lower balnicity index.
Our sample also has an unusually high fraction of LoBALs, four to
five out of seven, compared to the above samples. The most likely
explanation for the high frequency of BALs and LoBALs in our
sample is the red colour selection, $O-E \ge 2$. A comparison of
the three studies suggests that the BAL fraction in radio-selected
QSO samples is at least as high as in optically-selected ones.

(iv) We note the unusual QSO FIRST 1413+4505
($z=3.11$), whose optical spectrum reveals a strong associated
Ly$\alpha$ absorber (damped or BAL-like) that completely removes the
Ly$\alpha$ emission. The source is highly luminous in the radio,
has a flat
radio spectral index, and is detected in $IRAS$ 12 and 25 $\mu$m,
yielding a bolometric luminosity in the range from
SDSS $z$ band to 25 $\mu$m $L \sim 1.8 \times 10^{15} L_\odot$, which
places it amongst the most luminous objects known.

(v) Using a sub-sample of seven QSOs with $3.7 \le z \le 4.4$ and $E
\le 19.1$, we derive a space density of radio loud QSOs, $\rho [M_{\rm AB}
(1450) \le -26.6 , P_{\rm 1.4 \ GHz} \ge 10^{25.7} ({\rm W \
Hz^{-1}})] = 1.7 \pm 0.6$ Gpc$^{-3}$.
Assuming a radio-loud
fraction $13.4 \pm 3$ per cent, the space density for all QSOs is
$\rho [M_{\rm AB} (1450) \le -26.6] = 12.5 \pm 5.6$ Gpc$^{-3}$. This
result is in good agreement with the best-fitting model in Fan et
al. (2001a,b), considering the statistical errors.

\section*{Acknowledgments}

We thank the referee for rapid and helpful feedback. RC, JIGS,
CRB, and MV acknowledge financial support from the Spanish
Ministerio de Educaci\'on y Ciencia under project AYA 2002-03326.

This research is in part
based on observations made with the Italian Telescopio Nazionale
Galileo (TNG) operated on the island of La Palma by the Fundaci\'on
Galileo Galilei of the INAF (Istituto Nazionale di Astrofisica) at
the Spanish Observatorio del Roque de los Muchachos of the
Instituto de Astrofisica de Canarias. Funding for the creation and
distribution of the SDSS Archive has been provided by the Alfred
P. Sloan Foundation, the Participating Institutions, the National
Aeronautics and Space Administration, the National Science
Foundation, the U.S. Department of Energy, the Japanese
Monbukagakusho, and the Max Planck Society. The SDSS Web site is
http://www.sdss.org/. The SDSS is managed by the Astrophysical
Research Consortium (ARC) for the Participating Institutions. The
Participating Institutions are The University of Chicago,
Fermilab, the Institute for Advanced Study, the Japan
Participation Group, The Johns Hopkins University, the Korean
Scientist Group, Los Alamos National Laboratory, the
Max-Planck-Institute for Astronomy (MPIA), the
Max-Planck-Institute for Astrophysics (MPA), New Mexico State
University, University of Pittsburgh, University of Portsmouth,
Princeton University, the United States Naval Observatory, and the
University of Washington. This research has made use of the
NASA/IPAC Extragalactic Database (NED) which is operated by the
Jet Propulsion Laboratory, California Institute of Technology,
under contract with the National Aeronautics and Space
Administration. The 2MASS project is a collaboration between The
University of Massachusetts and the Infrared Processing and
Analysis Center (JPL/ Caltech). Funding is provided primarily by
NASA and the NSF. The University of Massachusetts constructed and
maintained the observatory facilities, and operated the survey.

\end{document}